\documentclass[graybox]{svmult}

\usepackage{mathptmx}       
\usepackage{helvet}         
\usepackage{courier}        
\usepackage{type1cm}        
\usepackage{makeidx}         
\usepackage{graphicx}        
\usepackage{multicol}        
\usepackage[bottom]{footmisc}

\makeindex             

\begin{document}

\title*{Concordance cosmology with particle creation}
\author{Saulo Carneiro}
\institute{Saulo Carneiro \at Instituto de F\'{\i}sica, Universidade Federal da Bahia, Salvador, BA, Brazil, \email{saulo.carneiro@pq.cnpq.br}.}
%
%
\maketitle

\abstract*{A constant-rate creation of dark particles in the late-time FLRW spacetime provides a cosmological model in accordance with precise observational tests. The matter creation backreaction implies in this context a vacuum energy density scaling linearly with the Hubble parameter, which is consistent with the vacuum expectation value of the QCD condensate in a low-energy expanding spacetime. Both the cosmological constant and coincidence problems are alleviated in this scenario. We discuss the cosmological model that arises in this context and present a joint analysis of observations of the first acoustic peak in the cosmic microwave background (CMB) anisotropy spectrum, the Hubble diagram for supernovas of type Ia (SNIa), the distance scale of baryonic acoustic oscillations (BAO) and the distribution of large scale structures (LSS). We show that a good concordance is obtained, albeit with a higher value of the present matter abundance than in the standard model.}

\abstract{A constant-rate creation of dark particles in the late-time FLRW spacetime provides a cosmological model in accordance with precise observational tests. The matter creation backreaction implies in this context a vacuum energy density scaling linearly with the Hubble parameter, which is consistent with the vacuum expectation value of the QCD condensate in a low-energy expanding spacetime. Both the cosmological constant and coincidence problems are alleviated in this scenario. We discuss the cosmological model that arises in this context and present a joint analysis of observations of the first acoustic peak in the cosmic microwave background (CMB) anisotropy spectrum, the Hubble diagram for supernovas of type Ia (SNIa), the distance scale of baryonic acoustic oscillations (BAO) and the distribution of large scale structures (LSS). We show that a good concordance is obtained, albeit with a higher value of the present matter abundance than in the standard model.}


${ }$

The gravitation of vacuum fluctuations is in general a difficult problem, since their energy density usually depends on the renormalization procedure and on an adequate definition of the vacuum state in the curved background. In the case of conformal fields in de Sitter spacetime, the renormalized vacuum density is $\Lambda \approx H^4$ \cite{Ford,Dowker,Davies,Starobinsky}, which in a low-energy universe leads to a too tiny cosmological term. 

In the case we consider the vacuum energy of interacting fields, it has been suggested that in a low energy, approximately de Sitter background the vacuum condensate originated from the QCD phase transition leads to $\Lambda \approx m^3 H$, where $m \approx 150$ MeV is the energy scale of the transition \cite{Schutzhold,Klinkhamer,Urban,Urban2,Urban3,Ohta,Holdom}. These results are in fact intuitive. In a de Sitter background the energy per observable degree of freedom is given by the temperature of the horizon, $E \approx H$. For a massless free field this energy is distributed in a volume $1/H^{3}$, leading to a density  $\Lambda \approx H^4$, as above. For a strongly interacting field in a low energy space-time, on the other hand, the occupied volume is $1/m^3$, owing to confinement, and the expected density is $\Lambda \approx m^3 H$.

Such a late-time variation law for the vacuum term can also be derived as a backreaction of the creation of non-relativistic dark particles in the expanding spacetime \cite{Alcaniz}. The Boltzmann equation for this process is
\begin{equation} \label{Boltzman}
\frac{1}{a^3}\frac{d}{dt}\left(a^3n\right)= \Gamma n,
\end{equation}
where $n$ is the particle number density and $\Gamma$ is a constant creation rate. By taking $\rho_m = nM$, it can also be written as
\begin{equation} \label{conservacao1}
\dot{\rho}_m + 3H\rho_m = \Gamma \rho_m,
\end{equation}
where $M$ is the mass of the created particle.

Let us take, in addition to (\ref{conservacao1}), the Friedmann equation
\begin{equation} \label{Friedmann}
\rho_m + \Lambda = 3H^2,
\end{equation}
with the vacuum term satisfying the equation of state $p_{\Lambda} = - \Lambda$. Using (\ref{conservacao1}) and (\ref{Friedmann}) we obtain the conservation equation for the total energy,
\begin{equation}
\dot{\rho} + 3H(\rho +p) = 0,
\end{equation}
provided we take 
\begin{equation}
\Lambda = 2\Gamma H + \lambda_0,
\end{equation}
 where $\lambda_0$ is a constant of integration\footnote[5]{Strictly speaking, this result is only exact if we neglect the conserved baryons in the balance equations. Since baryons represent only about $5\%$ of the total energy content, this can be considered a good approximation.}. Since there is no natural scale for this constant, let us make it zero. Then we have $\Lambda = 2\Gamma H$. This is the time-variation law predicted for the vacuum density of the QCD condensate, with $\Gamma \approx m^3$. Dividing it by $3H^2$, we obtain
\begin{equation} \label{winfried}
\Gamma = \frac{3}{2} \left(1-\Omega_m \right) H,
\end{equation}
where $\Omega_m = 1 - \Omega_{\Lambda} \equiv \rho_m/(3H^2)$ is the relative matter density (for simplicity, we are considering only the spatially flat case). 

In the de Sitter limit ($\Omega_m = 0$), we have $\Gamma = 3H/2$, that is, the creation rate is equal (apart from a numerical factor) to the thermal bath temperature predicted by Gibbons and Hawking in the de Sitter spacetime \cite{Gibbons}. It also means that the scale of the future de Sitter horizon is determined, through $\Gamma$, by the energy scale of the QCD phase transition, the last cosmological transition we have. For the present time we have, from (\ref{winfried}) (with $\Omega_m \approx 1/3$), $H_0 \approx \Gamma \approx m^3$, and hence $\Lambda \approx m^6$, where $H_0$ is the current Hubble parameter. The former result is an expression of the Eddington-Dirac large number coincidence \cite{MenaMarugan}. The later - also known as Zeldovich's relation \cite{Bjorken} - gives the correct order of magnitude for $\Lambda$.

The corresponding cosmological model has a simple analytical solution, which reduces to the CDM model for early times and to a de Sitter universe for $t \rightarrow \infty$ \cite{Borges}. It has the same free parameters of the standard model and presents good concordance when tested against type Ia supernovas, baryonic acoustic oscillations, the position of the first peak of CMB and the matter power spectrum \cite{Alcaniz,Pigozzo,Carneiro,Carneiro2,Borges2,Zimdahl}. Furthermore, the coincidence problem is alleviated, because the matter density contrast is suppressed in the asymptotic future, owing to the matter production \cite{Alcaniz,Borges2}.

With $\Lambda = 2\Gamma H$ we obtain, from the Friedmann equations, the solution \cite{Borges,Pigozzo,Carneiro,Carneiro2}
\begin{equation}\label{Hgeral}
\frac{H}{H_0} \approx \left\{ \left[1 - \Omega_{m0} + \Omega_{m0} (1
+ z)^{3/2}\right]^2 + \Omega_{r0} (1+z)^4 \right\}^{1/2}\;,
\end{equation}
where $\Omega_{m0}$ is the present relative matter density, and we have added conserved radiation with present density parameter $\Omega_{r0}$. As discussed in \cite{Pigozzo,Carneiro,Carneiro2}, for non-zero $\Omega_{r0}$ the expression (\ref{Hgeral}) is an approximate solution, differing only $1\%$ from the exact one, since $\Omega_{r0} \approx 8 \times 10^{-5} \ll 1$. For $\Omega_{r0} = 0$, the solution (\ref{Hgeral}) is exact.

For early times we obtain $H^2(z) = H_0^2 \Omega_{r0} z^4$, and the radiation era is indistinguishable from the standard one. On the other hand, for high redshifts the matter density scales as $\rho_m(z) = 3H_0^2 \Omega_{m0}^2 z^3$. The extra factor $\Omega_{m0}$ - as compared to the $\Lambda$CDM model - is owing to the late-time process of matter production. In order to have nowadays the same amount of matter, we need less matter in the past. Or, in other words, if we have the same amount of matter in the past (say, at the time of matter-radiation equality), this will lead to more matter today. We can also see from (\ref{Hgeral}) that, in the asymptotic limit $z \rightarrow -1$, the solution tends to the de Sitter solution. Note that, like the $\Lambda$CDM model, the above model has only two free parameters, namely $\Omega_{m0}$ and $H_0$. On the other hand, it can not be reduced to the $\Lambda$CDM case
except for $z\rightarrow -1$. In this sense, it is falsifiable, that is, it may be ruled out by observations.

The Hubble function (\ref{Hgeral}) can be used to test the model against background observations like SNIa, BAO and the position of the first peak in the CMB spectrum \cite{Pigozzo,Carneiro,Carneiro2}. The analysis of the matter power spectrum was performed in \cite{Borges2}, where, for simplicity, baryons were not included and the cosmological term was not perturbed. In a subsequent publication a gauge-invariant analysis, explicitly considering the presence of late-time non-adiabatic perturbations, has shown that the vacuum perturbations are indeed negligible, except for scales near the horizon \cite{Zimdahl}.

We show in Table I the best-fit results for $\Omega_{m0}$ (with $H_0$ marginalized) with three samples of supernovas: the SDSS and Constitution compilations calibrated with the MLCS2k2 fitter, and the Union2 sample. For the sake of comparison, we also show the best-fit results for the spatially flat $\Lambda$CDM model. We should have in mind that the Union2 dataset is calibrated with the Salt2 fitter, which makes use of a fiducial $\Lambda$CDM model for including high-$z$ supernovas in the calibration. Therefore, that sample is not model-independent and, in the case of the standard model, the test should be viewed as rather a test of consistence. From the table we can see that for the model with particle creation the concordance is quite good. For the samples calibrated with the MLCS2k2 fitter it is actually better than in the $\Lambda$CDM case. As antecipated above, the present matter density is higher than in the standard case. 

With the concordance values of $\Omega_{m0}$ in hand, we can obtain the age parameter of the Universe. It is given by \cite{Borges,Pigozzo,Carneiro,Carneiro2}
\begin{equation} \label{idade}
H_0t_0 = \frac{2\ln\Omega_{m0}}{3(\Omega_{m0}-1)} .
\end{equation}
In the case of the SDSS and Constitution samples, this leads to $H_0 t_0 = 0.97$, in good agreement with standard predictions and astronomical limits. For $H_0 \approx 70$ Km/(s.Mpc), we have $t_0 \approx 13.5$ Gyr.

\begin{table}[t]
\caption{$2\sigma$ limits to $\Omega_{m0}$ (SNe+ CMB + BAO+LSS).}
\begin{center}
\begin{tabular}{rcccc}
\hline \hline \\
\multicolumn{1}{c}{ } & \multicolumn{2}{c}{$\Lambda(t)$CDM } & \multicolumn{2}{c}{$\Lambda$CDM } \\
\multicolumn{1}{c}{Test}&
\multicolumn{1}{c}{$\Omega_{m0}$}&
\multicolumn{1}{c}{$\chi^2_{min}/\nu$}&
\multicolumn{1}{c}{$\Omega_{m0}$$^a$}&
\multicolumn{1}{c}{$\chi^2_{min}/\nu$}\\ \hline \\
Union2 (SALT2)&$0.420^{+0.009}_{-0.010}$ & 1.063 & $0.235\pm 0.011$ & 1.027 \\
SDSS (MLCS2k2)& $0.450^{+0.014}_{-0.010}$ & 0.842 & $0.260^{+0.013}_{-0.016}$ & 1.231 \\
Constitution (MLCS2k2-17)& $0.450^{+0.008}_{-0.014} $ &1.057 & $0.270\pm 0.013$ & 1.384\\
\hline \hline
\end{tabular}
\end{center}
\end{table}


${ }$

This work was also presented in {\it Relativity and Gravitation: 100 years after Einstein in Prague} (Prague, June 2012). It is dedicated to the memory of Prof. Pedro F\'elix Gonz\'alez-D\'{\i}az.

\bibliographystyle{spphys}
\bibliography{carneiro}

\end{document}